\documentclass[twocolumn]{aastex62}
\pdfoutput=1 %for arXiv submission
\usepackage{amsmath,amstext}
\usepackage[T1]{fontenc}
\usepackage{apjfonts} 
\usepackage[figure,figure*]{hypcap}

\usepackage{fnpos}

\shorttitle{Critical stellar rotation condition for the formation of a BH without bright transients}
\shortauthors{Murguia-Berthier et al.}

\begin{document}
\title{On the maximum stellar rotation to form a black hole without an accompanying luminous transient}
\author{Ariadna~Murguia-Berthier}
\affiliation{Department of Astronomy and Astrophysics, University of California, Santa Cruz, CA 95064, USA}
\affiliation{DARK, Niels Bohr Institute, University of Copenhagen, Blegdamsvej 17, 2100 Copenhagen, Denmark}
\author{Aldo~Batta}
\affiliation{Instituto Nacional de Astrof\'isica, \'Optica y Electr\'onica, Tonantzintla, Puebla 72840, M\'exico}
\affiliation{Consejo Nacional de Ciencia y Tecnolog\'ia, Av. Insurgentes Sur 1582, Col. Cr\'edito Constructor, CDMX, C.P. 03940, Mexico}
\affiliation{Department of Astronomy and Astrophysics, University of California, Santa Cruz, CA 95064, USA}
\affiliation{DARK, Niels Bohr Institute, University of Copenhagen, Blegdamsvej 17, 2100 Copenhagen, Denmark}
\author{Agnieszka~Janiuk}
\affiliation{Centrum Fizyki Teoretycznej PAN
Al. Lotnik\'ow 32/46, 02-668 Warsaw, Poland}
\affiliation{DARK, Niels Bohr Institute, University of Copenhagen, Blegdamsvej 17, 2100 Copenhagen, Denmark}
\author{Enrico~Ramirez-Ruiz}
\affiliation{Department of Astronomy and Astrophysics, University of California, Santa Cruz, CA 95064, USA}
\affiliation{DARK, Niels Bohr Institute, University of Copenhagen, Blegdamsvej 17, 2100 Copenhagen, Denmark}
\author{Ilya~Mandel}
\affiliation{Monash Centre for Astrophysics, School of Physics and Astronomy, Monash University, Clayton, Victoria 3800, Australia}
\affiliation{The ARC Center of Excellence for Gravitational Wave Discovery -- OzGrav}
\affiliation{Institute of Gravitational Wave Astronomy and School of Physics and Astronomy, University of Birmingham, Birmingham, B15 2TT, United Kingdom}
\affiliation{DARK, Niels Bohr Institute, University of Copenhagen, Blegdamsvej 17, 2100 Copenhagen, Denmark}
\author{Scott C.~Noble}
\affiliation{Gravitational Astrophysics Laboratory, NASA Goddard Space Flight Center, Greenbelt, MD 20771, USA}
\author{Rosa Wallace~Everson}
\affiliation{Department of Astronomy and Astrophysics, University of California, Santa Cruz, CA 95064, USA}
\affiliation{DARK, Niels Bohr Institute, University of Copenhagen, Blegdamsvej 17, 2100 Copenhagen, Denmark}

\begin{abstract}
The collapse of a massive star with low angular momentum content is commonly argued to result in the formation of a black hole without an accompanying  bright transient. Our  goal in this Letter is  to  understand  the  flow in and around a newly-formed black hole, involving accretion and  rotation,  via  general relativistic hydrodynamics simulations  aimed at studying the conditions under which infalling material can accrete without forming a centrifugally supported structure and, as a result,  generate no effective feedback. 
If the feedback from the black hole is, on the other hand, significant, the collapse would be halted and  we suggest that the event is likely to be followed by a bright transient.  We find that feedback is only efficient if the specific angular momentum of the infalling material at the innermost stable circular orbit exceeds that of geodesic circular flow at that radius by at least $\approx 20\%$. We use the results of our simulations to constrain the maximal stellar rotation rates of the disappearing massive progenitors  PHL293B-LBV and  N6946-BH1, and to provide an estimate of  the overall rate of
disappearing  massive stars. We find that about  a few percent  of single O-type stars with measured rotational velocities  are expected to spin below the critical value before collapse and are thus predicted to vanish without a trace.
\end{abstract}
\section{Introduction}
Recent evidence for  the disappearance of massive stars  \citep{2015MNRAS.450.3289G,2017MNRAS.468.4968A,allan2020disappearance} emphasizes the importance of studying the formation of black holes (BHs) and the conditions under which their formation might  trigger  a bright transient event \citep{1999ApJ...522..413F,2006ApJ...637..914W,2013ApJ...769..109L,2015PASA...32...16S,2015MNRAS.446.1213K,2016ApJ...821...38S}.

It is widely believed  that the lack of a bright transient is due to the collapse of a slowly rotating star \citep{1999ApJ...522..413F,2015PASA...32...16S}.  In this scenario, it is commonly assumed that the central engine involves a newly-formed BH accreting  material from the collapsing star. 
The properties of the inflowing material depend on the internal structure 
of the pre-collapse star and,  in particular,  its angular momentum  \citep{2014ApJ...781..119P, Lee06,Zalamea09}. 
The angular momentum content of the stellar progenitor is a key ingredient as  even a small amount of  rotation can break spherical symmetry and could produce a centrifugally-supported accretion disk,  which will evolve via internal magneto-hydrodynamic (MHD) stresses  \citep{1991ApJ...376..214B}. It has been noted that even in the absence of rotation, convective motions in the outer parts of highly evolved stars could also produce accretion disks \citep{2014MNRAS.439.4011G, 2016ApJ...827...40G, 2019MNRAS.485L..83Q}.

Spherical accretion onto BHs is relatively inefficient 
at producing feedback because the material is compressed but not shocked and thus cannot effectively convert gravitational to thermal energy \citep{1952MNRAS.112..195B,2012ApJ...752...30B}.
This changes dramatically when the infalling material has a critical amount of specific angular momentum \citep{1988ApJ...335..862F,2015ApJ...803...41M}.
When this is the case and if material is injected at large radii, a standard  accretion disk will form. Disk material will then gradually spiral inwards as internal MHD stress transports its angular momentum outwards.  

Accretion disks naturally produce MHD winds, which carry both bulk kinetic energy and ordered Poynting flux \citep{2011MNRAS.418L..79T,2012MNRAS.423.3083M}. The energy released by this accretion disk feedback is expected to be significantly larger than the binding energy of the star 
\citep{2005ApJ...629..341K,2014ARA&A..52..529Y}, which implies that the   motion  of the inflowing stellar gas can be effectively reversed. If the inflow is halted, we can then set constraints on the final mass and spin of the newly-formed BH \citep{2019arXiv190404835B}.  Our understanding of the fate of the collapsing star thus depends  on  our ability to determine the critical specific angular momentum below which material is able to accrete without generating feedback. 

General  relativity   plays  a  crucial  role  and sets the specific angular momentum 
at the innermost stable circular orbit. The flow pattern changes dramatically if the specific angular momentum of the inflowing material is near this critical value, as gas  will not only be compressed but will be able to dissipate its  motion perpendicular to the plane of symmetry and form a disk that is only marginally supported  by rotation \citep{Beloborodov01,Lee06,Zalamea09}. As the specific angular momentum increases, the rotational support becomes progressively more dominant until a standard Keplerian disk is formed. In this {\it Letter} we perform the first multidimensional general relativistic  simulations of uniformly rotating, low angular momentum  non-magnetized  flows  (Section~\ref{methods}), in order to derive the properties of the flow near this critical transition  (Section~\ref{results}) and establish when feedback becomes relevant (Section~\ref{energy}). We then make use of these results  to obtain an upper limit on the angular momentum that would allow the observed massive stellar progenitors to vanish without a trace (Section~\ref{dis}).

\section{Numerical setup and Initial Conditions}\label{methods}

We performed two-dimensional numerical simulations of low angular momentum, flows using the Eulerian code HARM \citep{Gammie03,Noble06}, which solves the equations of general relativistic MHD (GRMHD).  
Our setup consists of a quasi-radial inflow of  non-magnetized gas onto an accreting BH. The infalling gas has  specific angular momentum near the critical value, defined as that assigned to the innermost stable circular orbit (ISCO) of a BH. The numerical setup is similar to the one described in  \citet{2015MNRAS.447.1565S}, \citet{2017MNRAS.472.4327S}, \citet{2018ApJ...868...68J} and \citet{2019MNRAS.487..755P}.

 The boundary conditions in the angular direction are set to be periodic while the outer inner boundary is set to be out-flowing and the outer radial boundary is set to the inflow condition.
 This boundary is placed at large enough radii such that it will not impact the central region over the duration of the simulation  ($\approx 300 r_{\rm g}/c$) \citep{2017MNRAS.472.4327S}. 
 
 The units of the code are in the geometric system in which lengths are expressed in terms of the gravitational radius
 \begin{equation}
 r_{\rm g}=\frac{GM_{\rm bh}}{c^2},
 \end{equation}
where $M_{\rm bh}$ is the mass of the BH.  For converting to cgs units, we used the same convention as that described in \citet{2019ApJ...882..163J}.  In this convention, if $M_{\rm bh}=1M_\odot$, the time unit is $5\times 10^{-6}$s and  $r_{\rm g}=1.48$km. In our particular case, we choose $M_{\rm bh}=20M_\odot$, which corresponds to a time unit of $9.9\times 10^{-5}$s, and a length unit of $29.5$km. For our simulations,  the enclosed mass in the computational domain, defined as $2\pi \int^\pi_0 \int^{R_{\rm domain}}_{R_{
\rm in}} \rho \sqrt{-g} dr d\theta $, is chosen to be $0.2M_{\odot}$ (where $g$ is the determinant of the metric, $R_{\rm in}$ is the inner radius, and $R_{\rm domain}$ is the domain size), which in turn corresponds to a mass accretion rate of $0.1M_{\odot}$/s.

The domain covers $R_{\rm domain}=200 r_{\rm g}$ around the BH for simulations with a non-spinning BH, and $R_{\rm domain}=100 r_{\rm g}$ for simulations with spin. The resolution is $800\times800$ cells in the $x_1$ and $x_2$ directions, where $x_1$ and $x_2$ are the coordinates in spherical Kerr-Schild form for a non-spinning BH, and $400\times400$ for a BH with spin.
The initial radial component of the velocity ($u^r$) of the material is determined by the relativistic version of the Bernoulli equation \citep{1986bhwd.book.....S}.  In this formalism, the critical point ($r_{\rm s}$,  where subscript $s$ stands for the sonic point), where the flow becomes supersonic, is set as a free parameter. In this case, the critical point lies outside the domain at $r_{\rm s}=1000r_{\rm g}$, resembling a collapsing $34M_{\odot}$ star from models of \citet{2006ApJ...637..914W}. This implies that matter is always supersonic within our computational domain. The fluid is considered a polytrope with a pressure $P=K\rho^\gamma$, where $\rho$ is the density,  $\gamma=4/3$ is the adiabatic index, and $K$ is the specific entropy, in this case taken to be that of a relativistic fluid with inefficient cooling.  In what follows we describe how we generate the initial conditions.

Once the critical point is determined, the velocity at this critical point is \citep{1986bhwd.book.....S}:
\begin{equation}
 [u^r_{\rm s}]^2=\frac{GM_{\rm bh}}{2r_{\rm s}}, 
\end{equation}
 where $r$ is the radial coordinate and  $u^r$ is the radial component of the four-velocity. 
The radial velocity can be obtained by numerically solving the relativistic Bernoulli equation:
\begin{equation}
\bigg{(} 1+\frac{\gamma}{\gamma-1}\frac{P}{\rho}\bigg{)} ^2 \bigg{(}  1-\frac{2GM_{\rm bh}}{r}+[u^r]^2 \bigg{)} = \rm{constant},
\end{equation}
and the density is set by the mass accretion rate $\dot{M}$: 
\begin{equation}
\rho={\frac{\dot{M}}{4\pi r^2u^r}}.
\end{equation}

The specific entropy value, $K$, depends on the radial velocity and is taken to be \citep{2015MNRAS.447.1565S,2017MNRAS.472.4327S,2019MNRAS.487..755P}:
\begin{equation}
K=\bigg{(} u^r 4\pi r^2\frac{c_{\rm s}^{\frac{2}{\gamma-1}}}{\gamma^{\frac{1}{\gamma-1}}\dot{M}}\bigg{)}^{\gamma-1},
\end{equation}
where $c_{\rm s}^2=\frac{\gamma P}{\rho}$ is the local sound speed.

\begin{figure*}[t!]
\includegraphics[scale=0.416]{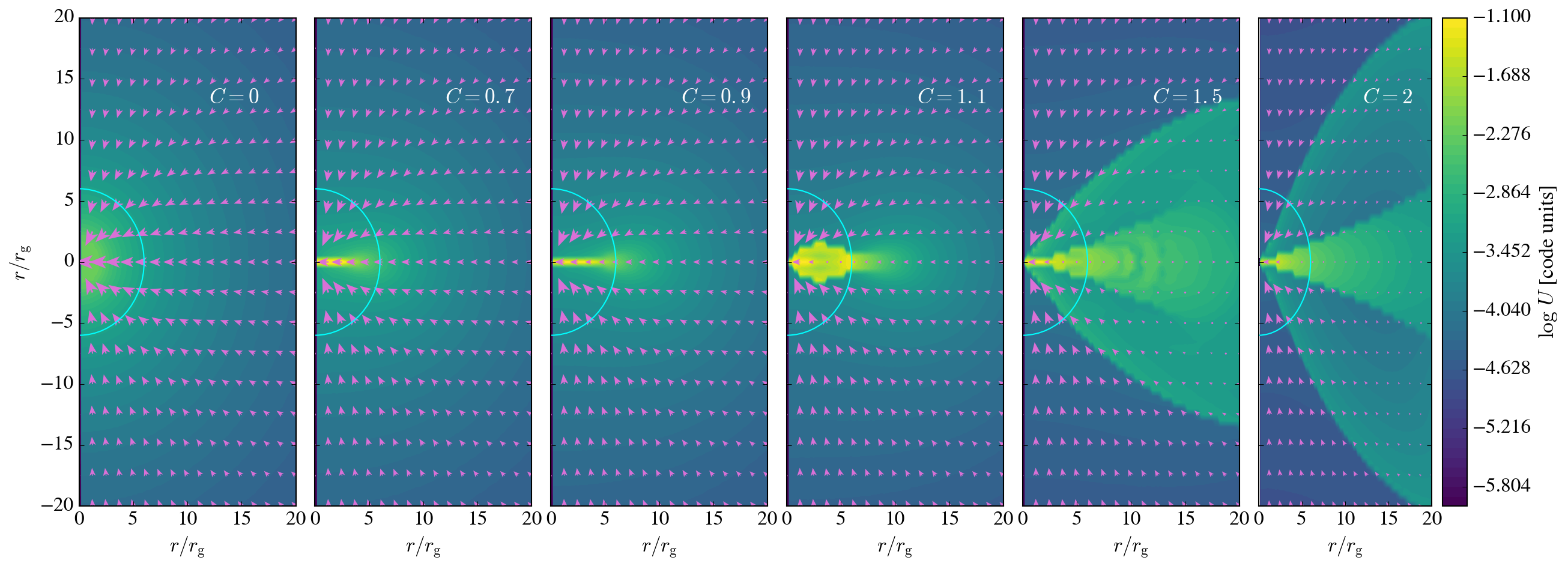}
\caption{Contour plot of internal energy density (in code units) at $t=300 r_{\rm g}/c$ for simulations of initially non-spinning BHs ($a_0=0$) with varying $C$. The arrows represent the velocity vectors of the flow, and the cyan circle shows the location of the ISCO. } 
\label{fig:same_time}
\end{figure*}

\begin{figure}[t!]
\includegraphics[scale=0.25]{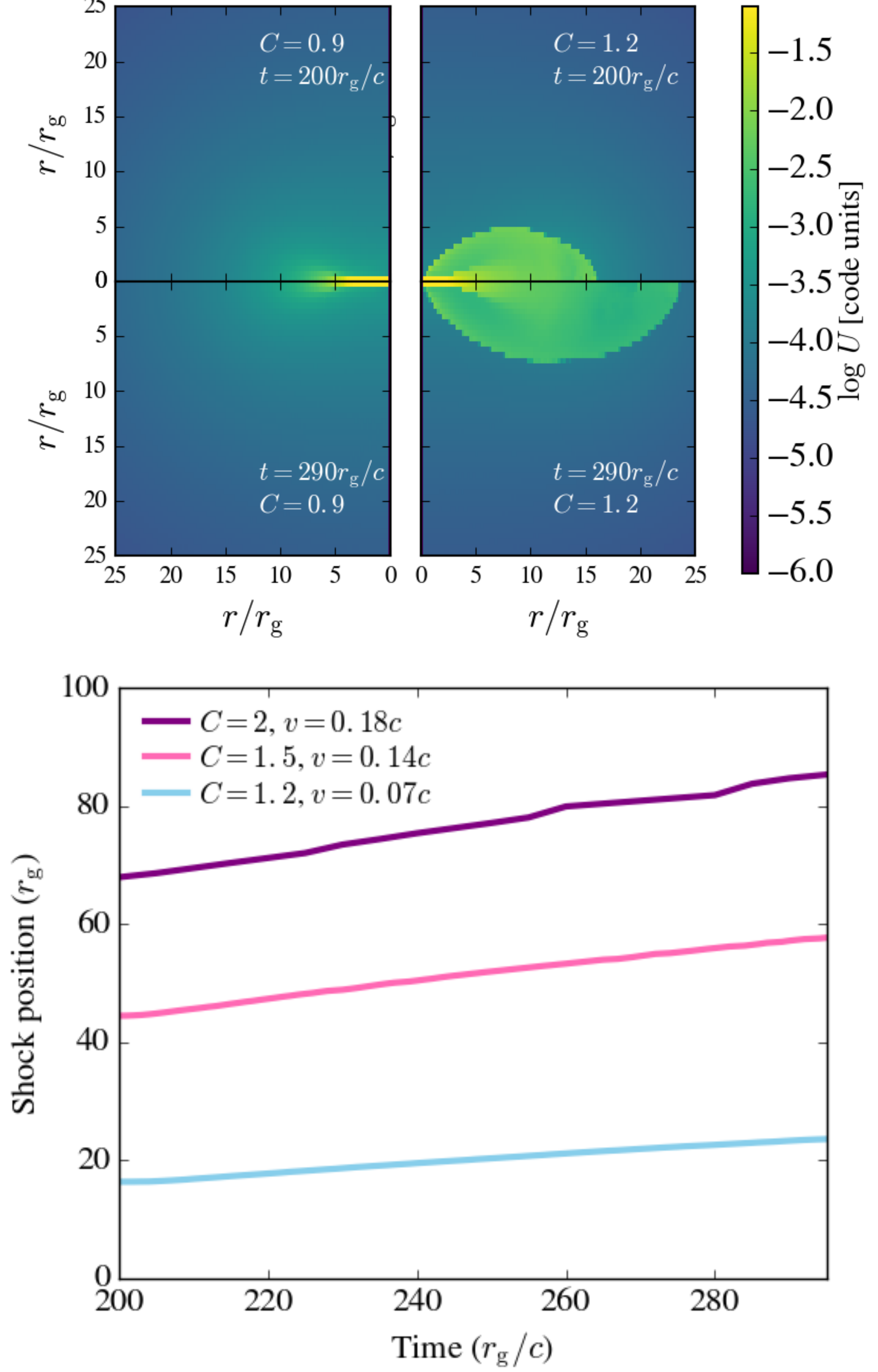}
\caption{\emph{Top} panel: The evolution  of the  internal energy density for two different simulations with  $C=0.9$ and $C=1.2$ plotted  at two different times for BHs with initial spin $a_0=0$. The resolution is the same as in Figure~\ref{fig:same_time}.
 \emph{Bottom} panel: The location of the shock discontinuity in the equatorial plane as a function of time. Plotted here are the shock locations for $C=2$ (purple line), $C=1.5$ (pink line), and $C=1.2$ (blue line). Shown in the legend are the  average shock front expansion velocities measured at the equator for the different values of $C$.  }
\label{fig:different_time}
\end{figure}

In order to derive the angular velocity at each radius, we use the  specific energy  and  angular momentum at the ISCO \citep{2015MNRAS.447.1565S,2017MNRAS.472.4327S,2019MNRAS.487..755P}:
\begin{equation}
\epsilon_{\rm isco}=-u_{t, \rm isco}=\frac{1-2/r_{\rm isco}+a/r_{\rm isco}^{3/2}}{\sqrt{1-3/r_{\rm isco}+2a/r_{\rm isco}^{3/2}}}
\end{equation}
and
\begin{equation}
l_{\rm isco}=u_{\phi, \rm isco}=\frac{r_{\rm isco}^{1/2}-2a/r_{\rm isco}+a^2/r_{\rm isco}^{3/2}}{\sqrt{1-3/r_{\rm isco}+2a/r_{\rm isco}^{3/2}}},
\end{equation}
where the radius of the ISCO $r_{\rm isco}$ in units of $r_{\rm g}$ if a function of the dimensionless BH spin $a$.   The angular velocity in Boyer-Lindquist coordinates for a Kerr metric can then be constructed as
\begin{equation}
u^\phi=g^{\phi \nu}u_\nu,
\end{equation}
where $\nu$ is an index used for Einstein summation notation, $\nu$ belongs to $\{t,r,\theta,\phi\}$.
For geodesic circular motion at the ISCO, the angular velocity is thus
\begin{equation}
u^\phi_{\rm isco}=-g^{\phi t}\epsilon_{\rm isco}+g^{\phi\phi}l_{\rm isco},
\end{equation}
where the components of the Kerr BH metric are $g^{t \phi}=-2ar/(\Sigma \Delta)$ and $g^{\phi \phi}=(\Delta-a^2\sin^2\theta)/(\Sigma \Delta \sin^2\theta)$, with $\Sigma=r^2+a^2\cos^2\theta$, $\Delta=r^2-2r+a^2$, and $\theta$ is the angular coordinate. 

In our simulations, we include a factor $C\sin^2\theta$ in the initial angular velocity profile such that
\begin{equation}
u^\phi=C \sin^2\theta (-g^{t\phi}\epsilon_{\rm isco}+g^{\phi\phi}l_{\rm isco}).
\end{equation}
The factor $\sin^2 \theta$ ensures that the angular momentum vanishes smoothly in the polar regions \citep{2017MNRAS.472.4327S}, and $C$ is a parameter that we vary.  Note that $C=0$ corresponds to  Bondi spherical accretion.

The initial angular momentum per unit mass is then given by $l=u_\phi=g_{\phi \nu}u^\nu$. In the case of $a=0$, it reduces to
\begin{equation}
l=Cl_{\rm isco} \sin^2\theta.
\end{equation}

In what follows we study the outcome  of our simulations as we systematically vary $C$ from the classical $C=0$ (spherical Bondi) to $C=2$.  This allows us to study the formation of accretion disks in low angular momentum flows along with exploring the dissipation of energy in the flow and ensuing feedback.

\section{Low angular momentum flows}\label{results}
As the star collapses, material will flow towards the newly formed BH and its angular momentum content will determine the final fate of the accreting object. If there is even a small amount of angular momentum, there will be dissipation of energy at the equator as material is shocked rather than solely compressed \citep{Beloborodov01,Lee06,Zalamea09}.

If the specific angular momentum is below critical, the energy dissipation  will be small and the heated gas will be promptly advected onto the BH. This is shown in Figure~\ref{fig:same_time}, where we plot contours of internal energy density and velocity vectors from  simulations with varying  $C$. The internal energy density in our simulations is related to the pressure as $U=\frac{P}{\gamma-1}$. As the specific angular momentum increases, material will be marginally bound and shocked near the equator before being accreted. When the angular momentum is near the critical one,  a shock discontinuity forms that steadily dissipates energy, which leads to a significant pressure build up. This is most evidently seen in the simulations at  around $C=1.1$.   This pressure build up  slows down the incoming material and produces  an angular momentum redistribution shock. It is noteworthy to point out that this shock is only transonic for the case of $C=2$. It is useful to compare the energy density in cases with higher angular momentum to the case $C=0$, where we expect inefficient feedback. 

As more material accumulates near the ISCO, the pressure supported structure  grows and expands for $C\gtrsim 1.2$, ultimately halting the flow.   The \emph{top} panel of Figure~\ref{fig:different_time} compares the time evolution of the energy dissipation for simulations with $C=0.9$ and $C=1.2$.  In the case of $C=0.9$, where the specific angular momentum is below the critical one, the dissipated energy is advected with the flow before being accreted by the BH. When $C=1.2$, a rotationally supported structure forms, which creates an expanding high-pressure region or {\it hot bubble}.  The energy accumulation in this region continues until the end of the simulations, leading to the steady increase of the bubble's size. This steady accumulation of energy could,  in principle, halt the collapse of the infalling star and  cause the envelope to be disrupted. The \emph{bottom} panel of Figure~\ref{fig:different_time} shows the position of the shock in the equatorial plane as a function of time, as well as the velocity of the shock. The shock moves outward with a velocity that is roughly constant in time and is larger than the escape velocity at the outer edge of the computational domain (which is $0.07c$). The material inside the shock will gain internal specific energy similar to the shock's kinetic energy, which is larger than the specific binding energy at the edge of the computational domain. This means that the expanding shock will be able to halt the collapse and effectively unbind the material at the edge of the computational domain.  However, this should be treated with caution, because it ignores the pressure from external material, which may act as a lid.  In order to reach firm conclusions about the fate of the collapsing star, we need to track the long-term evolution of the shock as it evolves through the entire stellar interior.

 We note that in our simulations, we don't include the effects of a changing metric, which are explored by \citet{2018ApJ...868...68J}. Not surprisingly, the authors  found that the BH accretes matter more rapidly for a changing metric, which can potentially alter the critical value of $C$. However, this effect is only relevant in our simulations at times that are much larger than those currently explored. This is becaus  throughout our simulation, the BH only accretes  a fraction $ \lesssim 0.01$ of its own mass, and thus the effects of both the self-gravity of the gas residing in the box  and the corresponding  change in the metric can be safely ignored.
 The critical angular momentum can also be altered by the inclusion of magnetic fields in the pre-collapse progenitor as well as the inclusion of radiation feedback. In the former case there can be additional outflows driven by the magnetic field stresses that can inject extra energy into the infalling material \citep{2012MNRAS.423.3083M,2019ApJ...882..163J,2019arXiv190404835B}. In the latter case, we expect that photons will be entirely advected onto the BH by the very optically thick accretion flow that is many orders of magnitude above the Eddington mass accretion limit in our simulation. The material is also expected to be optically thick to neutrinos, but if hypercritical accretion produces a neutrino-driven outflow, it could further help unbind the star \citep{2005ApJ...629..341K}.

\section{Energy Dissipation and Feedback}\label{energy}

\begin{figure}
\includegraphics[scale=0.6]{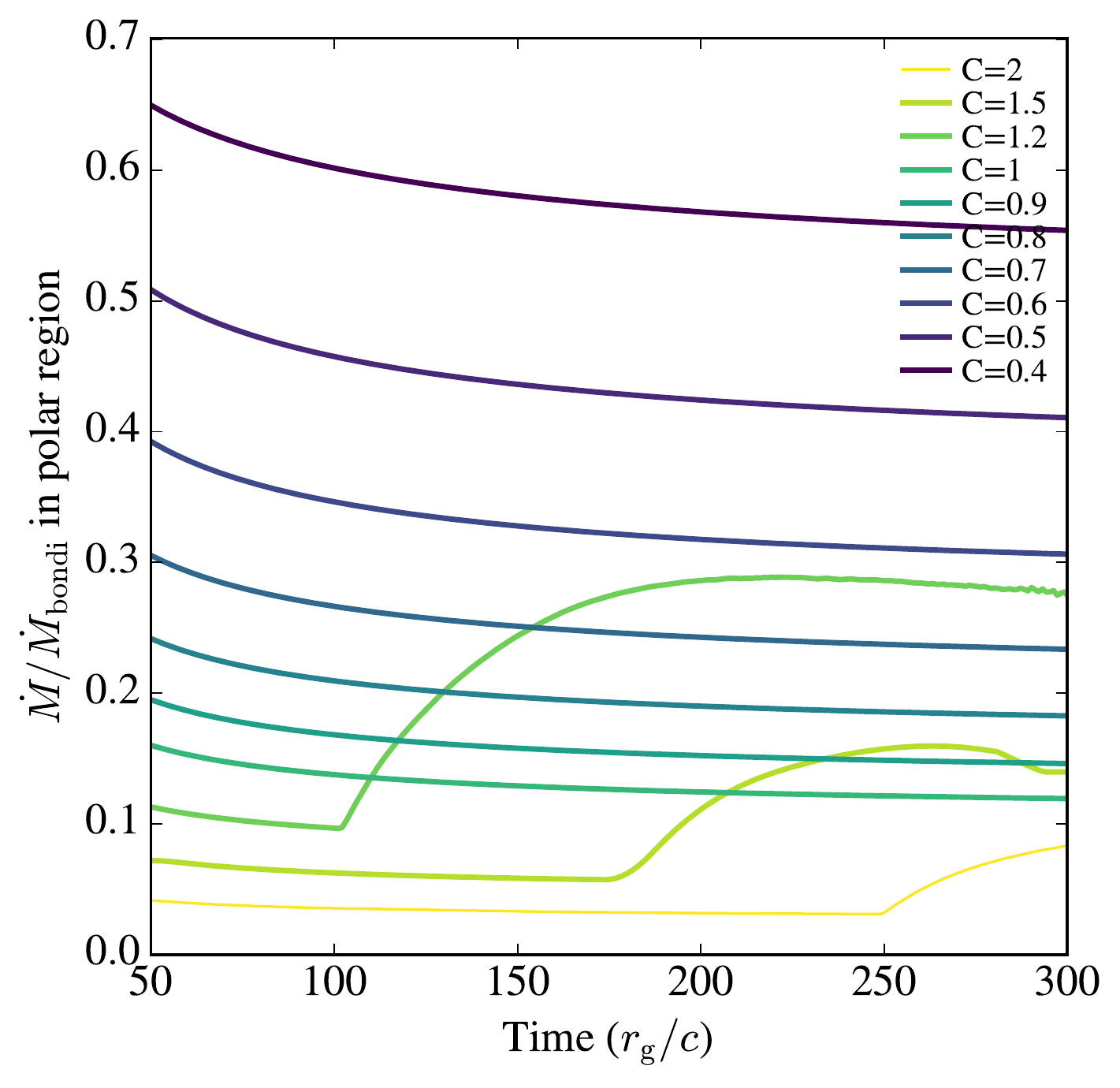}
\caption{Mass accretion rates (in units of $\dot{M}_{\rm bondi}$) in the polar region for simulations with varying $C$ and initially non-spinning BHs. The values for both $\dot{M}$ and $\dot{M_{\rm bondi}}$ are averaged at the ISCO over one quadrant of the simulation. The polar region is defined here by $0^\circ \leq \theta < 60^\circ$, with $\theta=90$ corresponding to the equatorial plane.} 
\label{fig:mdot}
\end{figure}

\begin{figure}
\includegraphics[scale=0.55]{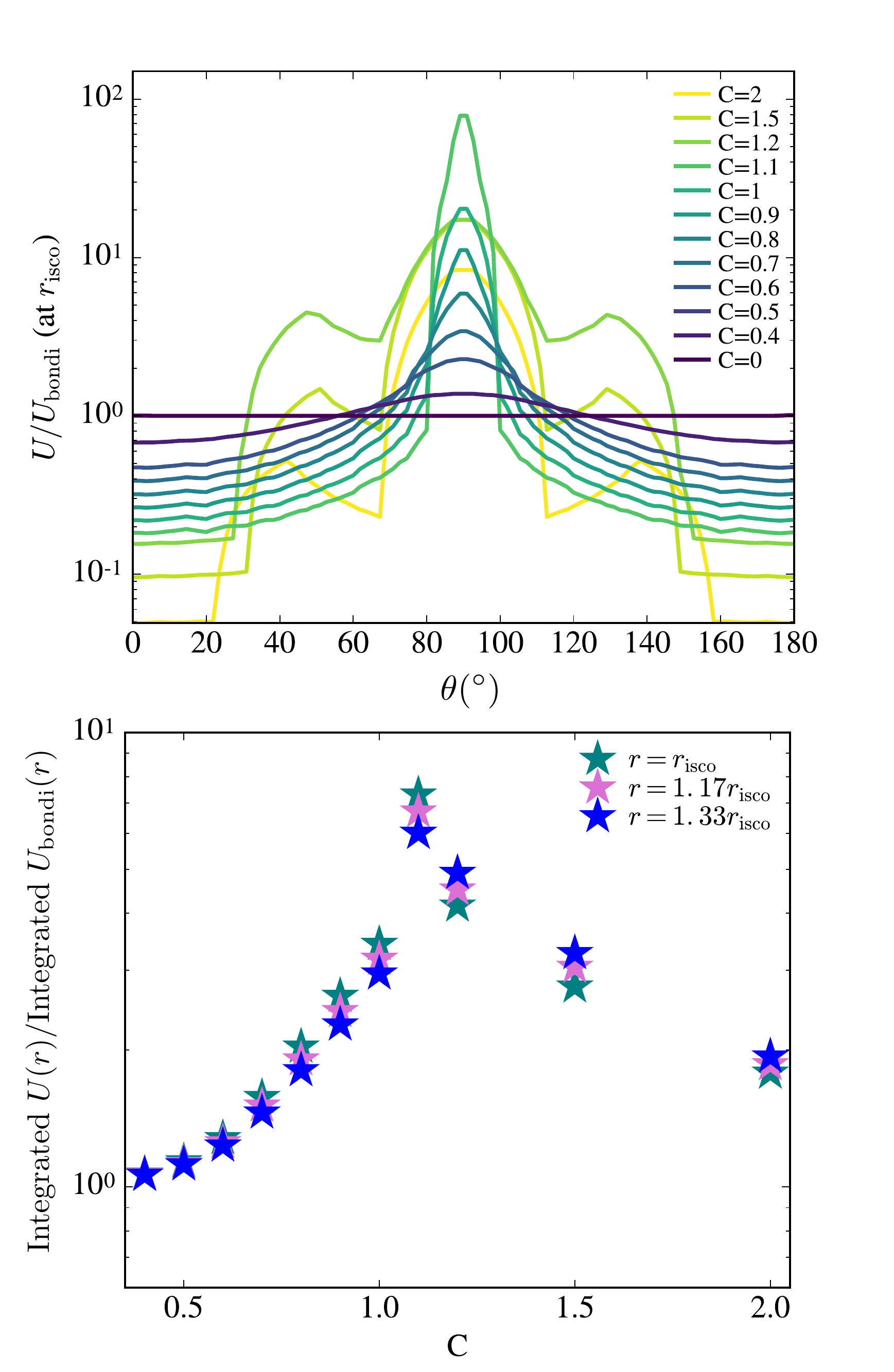}
\caption{The dissipation of energy in low angular momentum flows. \textit{Top} panel: Internal energy density at the ISCO as a function of $\theta$ for  initially non-spinning  BHs. 
Here  $\theta=90^\circ$ corresponds to the equator.  The normalization factor ($U_{\rm bondi}$) corresponds to  $C=0$ case, which is  spherically symmetric accretion and is solely driven by the compression of the flow.  \textit{Bottom} panel: Integrated internal energy out to a given radial scale 
as a function of  $C$. The integrated internal energy is calculated as  $U(r) = 2\pi\int^{\pi}_{0}\int^r_{2r_{\rm g}}\sqrt{-g(r')}U(r^\prime) dr^\prime d\theta$, 
where $g$ is the determinant of the metric and we use $r=[1, 1.17,1.33]r_{\rm isco}$. All the analyses make use of the snapshot at $t=300 r_{\rm g}/c$ for all simulations.}
\label{fig:energy_diss}
\end{figure}

As shown in Section~\ref{results}, the dissipation of energy in the infalling  gas  from   a collapsing star with $C\gtrsim 1.2$  can steadily accumulate near the equatorial plane. 
In this case, the energy dissipation rate exceeds the advection rate as the size of the dissipation region increases and, as a result, a hot pressure region or bubble is produced. This bubble, surrounded  by a clear discontinuity in both density and velocity, grows as material continues to be accreted. The corresponding  pressure build up halts the motion of the infalling material in the equatorial plane  while increasing the rate of accretion in the polar direction, as material at high latitudes is deflected towards the BH  (Figure~\ref{fig:same_time}). This can be seen in   Figure~\ref{fig:mdot}, which shows  the accretion rate in the polar direction as a function of time for all simulations with initial $a=0$ and  varying $C$.  

The amount of energy  dissipated  by accretion is commonly thought to be primarily determined by $\dot{M}$. Yet, since BHs do not have a hard surface, the feedback  efficiency  cannot be given solely by $\dot{M}$ as in the case of neutron stars or white dwarfs. Nor can BHs build up enough pressure to slow down the infalling gas.  Therefore, spherical accretion onto BHs advects any dissipated energy, without appreciable feedback. This situation changes dramatically when the inflow has a 
non-negligible amount of angular momentum and material is able to form a rotationally supported structure. In these cases, the energy dissipation rate is drastically altered. This can be seen  in Figure~\ref{fig:energy_diss}, where we plot in the top panel the internal energy density profile (normalized to Bondi) around the ISCO as a function of $\theta$. In this figure, $\theta=90^\circ$ corresponds to the equator and $\theta=0^\circ$ ($180^\circ$) to the polar direction.

Even though there is internal energy and mass accumulation when $C\lesssim 1.2$, feedback  will be inefficient because the flow is supersonic and the internal energy will be advected. The  dissipation rate increases dramatically  with $C$ as can be seen in the bottom panel of Figure~\ref{fig:energy_diss}. Plotted in this panel is the integrated energy density out to a given radial coordinate normalized  to the classical Bondi case ($C=0$).  The total dissipated energy increases as material with low angular momentum is shocked in the equatorial plane before being advected onto the BH.  A noticeable transition occurs at $C \approx 1.2$, as material begins to form a rotationally supported structure. The now differentially rotating  flow requires MHD stress in order to dissipate energy and transport  angular momentum, thereby enabling the inward accretion of gas. At this stage, the energy dissipation rate decreases as material becomes rotationally supported and shock dissipation is replaced by shear viscosity. In the absence of magnetic fields,  shear viscosity in our simulation is driven by  numerical dissipation,  which also acts over many orbital timescales. We thus caution the reader that the exact value of $C$  from our hydrodynamical simulations might be altered when internal MHD stresses are self-consistently included,  as a magnetized outflow can form that can further help halt the stellar collapse \citep{2012MNRAS.423.3083M,2019ApJ...882..163J,2019arXiv190404835B}. In our current simulations, it is around  $C \approx 1.2$ that we see the formation  of the  hot bubble, which continues to grow as the dissipated energy effectively accumulates near the ISCO (Figure~\ref{fig:different_time}).  
As the angular momentum continues to increase, a disk forms, which halts the advection of material and acts as a feedback term to slow the growth of energy dissipation near the ISCO.  
We thus conclude that for flows with $C\gtrsim 1.2$,  we expect feedback to likely halt the collapse of the infalling star. Because  the binding energy of failed SN progenitors steeply declines with increasing radius, it is suggested that any additional accumulation of energy will ultimately result in the disruption of the entire collapsing progenitor \citep{2019MNRAS.485L..83Q,2019arXiv190404835B}.
As the expanding envelope cools and radiation diffuses from it \citep[e.g.,][]{2020ApJ...892...13S}, a transient is expected  to accompany the formation of the BH \citep{1999ApJ...522..413F,2006ApJ...637..914W,2013ApJ...769..109L,2015PASA...32...16S,2015MNRAS.446.1213K,2016ApJ...821...38S,2019MNRAS.485L..83Q}. 

In addition to the initially non-spinning $a_0=0$ BH models, we also ran simulations with $a_0=0.05$ and $a_0=0.1$ and confirm that the feedback transition also occurs near  $C \approx 1.2$ and that the energy dissipation profiles are similar to  those plotted in Figure~\ref{fig:energy_diss}. This is consistent with \citet{2018ApJ...868...68J}, where the authors use a dynamical metric to explore how the accretion onto a BH influences the spin and final mass of the BH. They conclude that different initial spins lead to rather similar qualitative results, as we have found here.
\section{Discussion}\label{dis}
Having determined the critical specific angular momentum at which accretion onto a BH can  generate  feedback, we turn our attention to the conditions required for a stellar progenitor to collapse  without producing a bright transient under the assumption that significant feedback will unavoidably generate a discernible signal. In what follows, for simplicity, we assume that the star is uniformly rotating.

The corresponding critical angular velocity of the stellar progenitor is quantitatively estimated using the framework established by \citet{2019arXiv190404835B}, in which the formation and evolution of a BH is followed throughout the stellar collapse.  
For feedback not to be effective, the stellar progenitor needs to satisfy the following condition at all radii:
\begin{equation}
l(r)\leq l_{\rm fb} (r) = C_{\rm fb} l_{\rm isco}(r). 
\label{eq:limit}
\end{equation}
Here $C_{\rm fb}$ is the critical normalization factor taken to be $C_{\rm fb}=1.2$ and $l_{\rm isco}(r)$ is the specific angular momentum at the ISCO \citep{1972ApJ...178..347B}, which evolves as collapsing material is accreted by the BH.  

While rotating at such limiting angular velocity, only the star's outermost material has enough specific angular momentum $\Omega_{\rm lim} R_\ast^2$ to balance the critical condition $C_{\rm fb}\ l_{\rm isco}(R_\ast)$. At the same time, the rest of the material satisfies  condition~\ref{eq:limit}. In the ensuing subsections we express $\Omega_{\rm lim}$ in terms of the star's breakup angular velocity, $\Omega_{\rm break}=\left(G M_\ast/R_\ast^3\right)^{1/2}$, where $M_\ast$ and $R_\ast$ are the stellar mass and radius, respectively.

\begin{figure*}[t!]

\includegraphics[scale=0.5]{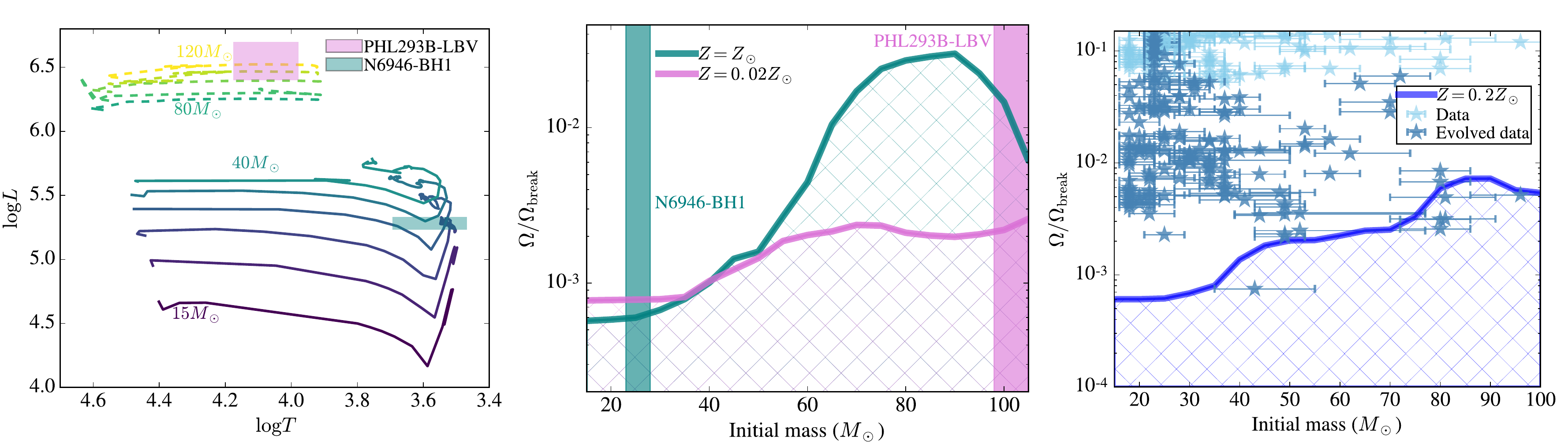}
\caption{\textit{Left} Panel: Hertzsprung-Russell diagram of the MESA models used in our analysis \citep{2011ApJS..192....3P,2013ApJS..208....4P}. The solid lines are models with  $Z=Z_\odot$.
The solid lines are stellar models that start at $M_{\rm ZAMS}=15M_\odot$ and are plotted every $5M_\odot$ until $40M_\odot$. In teal we show the luminosity and temperature constraints for N6946-BH1 \citep{2015MNRAS.450.3289G,2017MNRAS.468.4968A}. The dotted lines represent models with $Z=0.02Z_\odot$. Models start at $M_{\rm ZAMS}=80M_\odot$ and are plotted every $10M_\odot$ until $120M_\odot$. Models with $Z=0.02Z_\odot$ were used to constrain PHL293B-LBV \citep{allan2020disappearance}, whose luminosity and temperature constraints are shown in orchid. 
\textit{Middle} Panel: Maximum angular velocity at which a star can disappear without an accompanying  bright transient as a function of the initial mass of the progenitor. Here $\Omega_{\rm break}$ is the breakup velocity. 
The different lines are the constraints derived at different metallicities, which have been selected to match those of N6946-BH1 and PHL293B-LBV. Also plotted are the mass estimates we derive from our MESA models. Models in this specific  mass range spend a fraction of their last $10^4$ years of evolution within the corresponding uncertainty region in the HR diagram ({\it Left} Panel). 
\textit{Right} Panel: Angular velocity  as a function of the initial mass of the stellar progenitor. Plotted are the rotational velocities of single O-type  stars at $Z=0.2Z_\odot$ taken from \citet{2013A&A...560A..29R}, with masses derived by \citet{2010A&A...524A..98W}.  We evolve the rotational velocities of MESA models of these O-type stars by applying Equation \ref{eq:omegadot} and assuming rigid body rotation until carbon burning ends (see text for details about this assumption). These pre-collapse rotational velocities, labelled as {\it evolved data}, are compared with the range of angular velocities for these stars to collapse without an accompanying  bright transient ({\it hatched} region).
}
\label{fig:omega}
\end{figure*}

\subsection{On the disappearing stellar progenitors of  N6946-BH1 and PHL293B-LBV}
Let us now turn our attention to the properties of  N6946-BH1 and PHL293B-LBV, two stars that have been argued to disappear without an accompanying bright transient \citep{2015MNRAS.450.3289G,2017MNRAS.468.4968A,allan2020disappearance}. While other  explanations might be viable, a collapse to a BH without feedback is a possible explanation for the sudden disappearance of the star.

We use the stellar evolution code MESA \citep{2011ApJS..192....3P,2013ApJS..208....4P} version 8845 in order to constrain  the structure and observational properties of these stars. We use the default MESA parameters for massive stars.  For simplicity, our models are non-rotating, and their evolution is halted  when carbon burning ends. We ran the models using a Dutch hot wind scheme \citep{2009A&A...497..255G} with a scaling factor of 0.8. In this wind scheme, the mass loss rate prescription changes depending on the evolutionary stage of the star. For the rest of the paper, we take $Z_\odot=0.02$.
 
N6946-BH1 is a disappearing star found by \citet{2015MNRAS.450.3289G} and \citet{2017MNRAS.468.4968A} using  the \textit{Large Binocular Telescope}. The star is found to be embedded in a highly dusty environment in the galaxy NGC 6946. This red supergiant star was observed to increase its optical magnitude by around 5 magnitudes 
after a weak optical outburst in 2009. One possibility for this disappearing star is a collapse to a BH where the angular momentum was low enough that feedback  from the BH was unable to unbind the collapsing progenitor.

Information on the progenitor was deduced  using archival data from  the \textit{Hubble Space Telescope}, which was taken around two years before the weak outburst. Using dust and stellar evolution models, \citet{2015MNRAS.450.3289G} and \citet{2017MNRAS.468.4968A} deduced a luminosity of $\log L/L_\odot=5.29^{+0.04}_{-0.06}$ and a temperature of $T=3260^{+1670}_{-320}K$ for the pre-collapse progenitor. Their solar metallicity models constrained the progenitor mass to be $20-30M_\odot$.  

PHL293B-LBV \citep{allan2020disappearance} is another disappearing star. This luminous blue variable (LBV) was found in the
galaxy PHL293B.
\citet{allan2020disappearance} used ESO/VLT's ESPRESSO and X-shooter to obtain spectra of this galaxy in 2019. These spectra  lacked an LBV signature, which was  clearly present from 2011 to 2019. One of many viable possibilities is that when the eruptive period ended, the LBV collapsed into a BH.
Using radiative transfer models, \citet{allan2020disappearance} derived  a luminosity between $\log L/L_\odot= 6.3-6.7$ and a temperature between $T=9,500-15,000K$  for the pre-collapse star. %They use $Z=0.0004$ stellar models to  constrain its  mass to be somewhere between $85-120 M_\odot$.

We compare the temperature and luminosity constraints of N6946-BH1 and PHL293B-LBV with our stellar models in order to  constrain both their masses and internal structures. The {\it left} panel of figure \ref{fig:omega} shows the locations of N6946-BH1 and PHL293B-LBV on the Hertzsprung-Russell diagram together with the MESA stellar evolutionary models. The solid lines correspond to models with $Z=Z_\odot$, which are relevant to N6946-BH1 \citep{2015MNRAS.450.3289G,2017MNRAS.468.4968A}, while the dotted lines correspond to models with  $Z=0.02Z_\odot$, appropriate for  PHL293B-LBV \citep{allan2020disappearance}.  Using these models we constrain the initial masses of N6946-BH1  and PHL293B-LBV to be $23-28 M_\odot$ and $98-130 M_\odot$, respectively. These constraints are consistent  with those quoted in the literature. We caution the reader that given the mass range deduced  for  PHL293B-LBV, the final outcome could be a pair instability supernova \citep{2017ApJ...836..244W}. Nonetheless, the lack of a transient event for PHL293B-LBV suggests that this was not the case, as argued by \citet{allan2020disappearance}.

We use these models to also constrain the internal density structure of the progenitor, which in turn sets the moment of inertia and allows us to place a limit on the maximum angular velocity needed for the star to collapse without forming a disk. These limits for N6946-BH1  and PHL293B-LBV are plotted in the {\it middle} panel of Figure~\ref{fig:omega}. Within the  hatched region, the angular velocity of the  pre-collapse progenitor is  below the critical one in which feedback becomes efficient.  The regions extends to higher fractions of the break-up velocity for high-mass solar-metallicity stars because these stars self-strip due to rapid wind-driven mass loss, leaving behind compact, low moment of inertia Wolf-Rayet stars. 
We thus suggest that progenitors within this region will collapse without producing a bright transient.   

\subsection{Is it common for  stars to vanish without a trace?}
In the preceding sections we have endeavoured to outline the rotational constraints needed for  stellar progenitors to vanish without a trace. We caution that even in the absence of rotation, the outer layers might be still ejected by, for example, the loss of rest mass energy via neutrinos  \citep[e.g.,][]{2013ApJ...769..109L}
and could still produce a faint transient signal \citep[e.g.][]{2017ApJ...835..282M}.

Herein we assume that stellar spin is an essential parameter and turn to the problem of assembling the pre-collapse rotational constraints derived in this {\it Letter} into a general scheme involving the evolution of massive stars. In the {\it right} panel of figure~\ref{fig:omega} we plot the observationally derived rotation rates  of single O-type ($Z=0.2Z_\odot$) stars taken from \citet{2013A&A...560A..29R} with  initial stellar masses  derived by \citet{2010A&A...524A..98W}. 

We produce MESA models to match the age and stellar mass of these stars, using $Z=0.2Z_\odot$ and assuming rigid body rotation. Applying the observationally derived rotation rates, we then make use of the following standard relation \citep{1992MNRAS.257..450V}:
\begin{equation}\label{eq:omegadot}
\frac{1}{\Omega}\frac{d\Omega}{dt}=-\frac{1}{I_\ast}\frac{dI_\ast}{dt}+\frac{2}{3}\frac{R_\ast^2}{I_\ast}\frac{dM_\ast}{dt}.
\end{equation}
Where where $M_\ast$ and $R_\ast$ are the stellar mass and radius, respectively, $I_\ast=\frac{8\pi}{3}\int_0^{R_\ast} \rho(r) r^4dr$ is the moment of intertia of the star, and $\Omega$ is the angular velocity.
The evolution  of the rotational velocity is then computed until the end of the star's  life, which in our models corresponds to the end of carbon burning. In the  {\it right} panel of Figure~\ref{fig:omega}  we plot 
the final rotational velocity derived for each observed system  with the corresponding symbols labelled as {\it Evolved data}. 

Throughout this paper, we assumed rigid-body rotation, i.e., very efficient angular momentum transport within the star. It is evident that the mechanisms responsible for transporting angular momentum inside massive stars are currently 
not well understood \citep{2015ApJ...808...35K,2019ApJ...881L...1F}.  Even in the simplest case of uniform rotation, we find that stellar winds can extract a significant amount of angular momentum from
the star and in a small fraction of cases produce 
rotation rates close to those required for stars to vanish without a trace ({\it right} panel of Figure~\ref{fig:omega}). More specifically, we find that  $\approx$5\% of the stars we evolved (from a total of 163) have $(\Omega/\Omega_{\rm break})$ below the critical value ({\it hatched} region in the {\it right} panel of Figure~\ref{fig:omega}). In these cases we expect the collapse to proceed without the formation of an accretion disk, allowing the progenitor to vanish in our model. 

Although the evolution of O-type stars may be be commonly associated with supernovae, some of them might be expected to disappear.  If  single O-type stars with $(\Omega/\Omega_{\rm break})$ below the critical value are expected to vanish, we then conclude  that these objects are at least tens of  times rarer than standard supernova events. This of course has been derived under the assumption that a standard supernova event is the natural outcome for the vast majority of O-type stars with $(\Omega/\Omega_{\rm break})$ above the critical value.  Obviously, the above calculation is  limited and should be taken as an order of magnitude estimate at present. For example, using the same Dutch hot wind scheme in MESA but with a scaling factor of 1.0 (instead of the standard 0.8) we find that  $\approx$7\% of the stars we evolved  have $(\Omega/\Omega_{\rm break})$ below the critical (mass-dependent) value. 

This simple estimate for the rate of disappearing massive stars should improve as more objects have their rotational rates measured  and massive stellar evolution modelling improves. Having said this, it is important to note that this  few percent estimate is roughly  consistent with the one derived by \citet{2015MNRAS.450.3289G}, where they argued that the current rate of vanishing stars is $\gtrsim 7\%$ the  rate of core collapse supernova. This estimate can also be altered for red supergiants, as   convective motions in their outer layers might produce accretion disks and thus effective feedback even in the absence of net rotation  \citep{2019MNRAS.485L..83Q}.

Most massive stars are born in binaries, and binary interactions can significantly impact stellar structure and stellar rotation through mass transfer and tides \citep{Sana:2012}.  Accounting for the impact of binary evolution would further change the expected fraction of vanishing stars.

Many core collapses of massive stars are expected to produce supernovae when forming neutron stars in spherical explosions \citep{2012ASPC..453...91U,2016ApJ...821...38S} but some are expected  to have insufficient neutrino deposition \citep{1993ApJ...405..273W,2009ApJ...707..193F,2012ApJ...750...68L,2013ApJ...769..109L}
and will form a BH in the center of the star. 

The modeling of stellar collapse leading to BH formation is a formidable challenge  to computational techniques.  It is, also, a formidable challenge for observers, in their quest for finding   stars that disappear. If we were to venture on a general classification scheme for failed  supernovae, on the hypothesis that the central object involves a BH formed in a core collapse explosion, we expect the specific angular momentum of the infalling stellar material to be a critical parameter. When $l(r)\lesssim l_{\rm fb}(r)$  we predict the star will vanish without a trace. On the other hand, when $l(r)\gtrsim l_{\rm fb}(r)$  the collapse may instead be followed by a bright transient, whose properties will likely depend on the  mass and spin of the BH, the rate at which gas is supplied, the spin orientation relative to our line of sight, and the structure of the envelope through which any outflows will be re-processed.

\acknowledgements
We thank J. Schwab, R. Foley, A. Vigna-G\'omez, I. Palit, C. Kilpatrick, and the anonymous referee. We thank the Niels Bohr Institute, DARK Cosmology Centre in Copenhagen, University of Birmingham and Centrum Fizyki Teoretycznej for their hospitality while part of this work was completed. We also thank the Kavli Foundation for organizing the  Kavli Summer Program in 2017.  A.M.-B. acknowledges support from a UCMEXUS-CONACYT Doctoral Fellowship, and NASA TCAN award TCAN-80NSSC18K1488. A.B. was supported by the Danish National Research Foundation (project number DNRF132) and is currently funded by the program Cátedras CONACYT para Jóvenes Investigadores. E.R.-R. thanks the Heising-Simons Foundation, the Danish National Research Foundation (DNRF132) and NSF (AST-1911206 and AST-1852393) for support.  I.M. is a recipient of the Australian Research Council Future Fellowship FT190100574.
A.J. is supported in part by the Polish National Science Center under the grant DEC-2016/23/B/ST9/03114. S.C.N. was supported by NSF awards AST-1515982 and
OAC-1515969, NASA TCAN award TCAN-80NSSC18K1488, and by an appointment to the NASA Postdoctoral Program at
the Goddard Space Flight Center administrated by USRA through a
contract with NASA. R.W.E. is supported by the  National Science Foundation Graduate Research Fellowship Program (Award \#1339067), the Eugene V. Cota-Robles Fellowship, and the Heising-Simons Foundation. Any opinions, findings, and conclusions or recommendations expressed in this material are those of the authors and do not necessarily reflect the views of the National Science Foundation. The authors acknowledge use of the lux supercomputer at UC Santa Cruz, funded by NSF MRI grant AST 1828315.

\bibliography{minidisk_revision.bib}

\begin{thebibliography}{}
\expandafter\ifx\csname natexlab\endcsname\relax\def\natexlab#1{#1}\fi

\bibitem[{{Adams} {et~al.}(2017){Adams}, {Kochanek}, {Gerke}, {Stanek}, \&
  {Dai}}]{2017MNRAS.468.4968A}
{Adams}, S.~M., {Kochanek}, C.~S., {Gerke}, J.~R., {Stanek}, K.~Z., \& {Dai},
  X. 2017, Monthly Notices of the Royal Astronomical Society, 468, 4968

\bibitem[{Allan {et~al.}(2020)Allan, Groh, Mehner, Smith, Boian, \&
  Farrell}]{allan2020disappearance}
Allan, A., Groh, J., Mehner, A., {et~al.} 2020, The disappearance of a massive
  star in the low metallicity galaxy PHL 293B, , , arXiv:2003.02242

\bibitem[{{Balbus} \& {Hawley}(1991)}]{1991ApJ...376..214B}
{Balbus}, S.~A., \& {Hawley}, J.~F. 1991, \apj, 376, 214

\bibitem[{{Bardeen} {et~al.}(1972){Bardeen}, {Press}, \&
  {Teukolsky}}]{1972ApJ...178..347B}
{Bardeen}, J.~M., {Press}, W.~H., \& {Teukolsky}, S.~A. 1972, \apj, 178, 347

\bibitem[{{Batta} \& {Ramirez-Ruiz}(2019)}]{2019arXiv190404835B}
{Batta}, A., \& {Ramirez-Ruiz}, E. 2019, arXiv e-prints, arXiv:1904.04835

\bibitem[{{Beloborodov} \& {Illarionov}(2001)}]{Beloborodov01}
{Beloborodov}, A.~M., \& {Illarionov}, A.~F. 2001, \mnras, 323, 167

\bibitem[{{Blondin} \& {Raymer}(2012)}]{2012ApJ...752...30B}
{Blondin}, J.~M., \& {Raymer}, E. 2012, \apj, 752, 30

\bibitem[{{Bondi}(1952)}]{1952MNRAS.112..195B}
{Bondi}, H. 1952, \mnras, 112, 195

\bibitem[{{Fryer}(1999)}]{1999ApJ...522..413F}
{Fryer}, C.~L. 1999, \apj, 522, 413

\bibitem[{{Fryer} {et~al.}(2009){Fryer}, {Brown}, {Bufano}, {Dahl}, {Fontes},
  {Frey}, {Holland}, {Hungerford}, {Immler}, {Mazzali}, {Milne}, {Scannapieco},
  {Weinberg}, \& {Young}}]{2009ApJ...707..193F}
{Fryer}, C.~L., {Brown}, P.~J., {Bufano}, F., {et~al.} 2009, \apj, 707, 193

\bibitem[{{Fryxell} \& {Taam}(1988)}]{1988ApJ...335..862F}
{Fryxell}, B.~A., \& {Taam}, R.~E. 1988, \apj, 335, 862

\bibitem[{{Fuller} \& {Ma}(2019)}]{2019ApJ...881L...1F}
{Fuller}, J., \& {Ma}, L. 2019, \apjl, 881, L1

\bibitem[{{Gammie} {et~al.}(2003){Gammie}, {McKinney}, \&
  {T{\'o}th}}]{Gammie03}
{Gammie}, C.~F., {McKinney}, J.~C., \& {T{\'o}th}, G. 2003, \apj, 589, 444

\bibitem[{{Gerke} {et~al.}(2015){Gerke}, {Kochanek}, \&
  {Stanek}}]{2015MNRAS.450.3289G}
{Gerke}, J.~R., {Kochanek}, C.~S., \& {Stanek}, K.~Z. 2015, Monthly Notices of
  the Royal Astronomical Society, 450, 3289

\bibitem[{{Gilkis} \& {Soker}(2014)}]{2014MNRAS.439.4011G}
{Gilkis}, A., \& {Soker}, N. 2014, \mnras, 439, 4011

\bibitem[{{Gilkis} \& {Soker}(2016)}]{2016ApJ...827...40G}
---. 2016, \apj, 827, 40

\bibitem[{{Glebbeek} {et~al.}(2009){Glebbeek}, {Gaburov}, {de Mink}, {Pols}, \&
  {Portegies Zwart}}]{2009A&A...497..255G}
{Glebbeek}, E., {Gaburov}, E., {de Mink}, S.~E., {Pols}, O.~R., \& {Portegies
  Zwart}, S.~F. 2009, \aap, 497, 255

\bibitem[{{Janiuk}(2019)}]{2019ApJ...882..163J}
{Janiuk}, A. 2019, \apj, 882, 163

\bibitem[{{Janiuk} {et~al.}(2018){Janiuk}, {Sukova}, \&
  {Palit}}]{2018ApJ...868...68J}
{Janiuk}, A., {Sukova}, P., \& {Palit}, I. 2018, \apj, 868, 68

\bibitem[{{Kissin} \& {Thompson}(2015)}]{2015ApJ...808...35K}
{Kissin}, Y., \& {Thompson}, C. 2015, \apj, 808, 35

\bibitem[{{Kochanek}(2015)}]{2015MNRAS.446.1213K}
{Kochanek}, C.~S. 2015, \mnras, 446, 1213

\bibitem[{{Kohri} {et~al.}(2005){Kohri}, {Narayan}, \&
  {Piran}}]{2005ApJ...629..341K}
{Kohri}, K., {Narayan}, R., \& {Piran}, T. 2005, \apj, 629, 341

\bibitem[{{Lazzati} {et~al.}(2012){Lazzati}, {Morsony}, {Blackwell}, \&
  {Begelman}}]{2012ApJ...750...68L}
{Lazzati}, D., {Morsony}, B.~J., {Blackwell}, C.~H., \& {Begelman}, M.~C. 2012,
  \apj, 750, 68

\bibitem[{{Lee} \& {Ramirez-Ruiz}(2006)}]{Lee06}
{Lee}, W.~H., \& {Ramirez-Ruiz}, E. 2006, \apj, 641, 961

\bibitem[{{Lovegrove} \& {Woosley}(2013)}]{2013ApJ...769..109L}
{Lovegrove}, E., \& {Woosley}, S.~E. 2013, \apj, 769, 109

\bibitem[{{MacLeod} {et~al.}(2017){MacLeod}, {Macias}, {Ramirez-Ruiz},
  {Grindlay}, {Batta}, \& {Montes}}]{2017ApJ...835..282M}
{MacLeod}, M., {Macias}, P., {Ramirez-Ruiz}, E., {et~al.} 2017, \apj, 835, 282

\bibitem[{{MacLeod} \& {Ramirez-Ruiz}(2015)}]{2015ApJ...803...41M}
{MacLeod}, M., \& {Ramirez-Ruiz}, E. 2015, \apj, 803, 41

\bibitem[{{McKinney} {et~al.}(2012){McKinney}, {Tchekhovskoy}, \& {Bland
  ford}}]{2012MNRAS.423.3083M}
{McKinney}, J.~C., {Tchekhovskoy}, A., \& {Bland ford}, R.~D. 2012, \mnras,
  423, 3083

\bibitem[{{Noble} {et~al.}(2006){Noble}, {Gammie}, {McKinney}, \& {Del
  Zanna}}]{Noble06}
{Noble}, S.~C., {Gammie}, C.~F., {McKinney}, J.~C., \& {Del Zanna}, L. 2006,
  \apj, 641, 626

\bibitem[{{Palit} {et~al.}(2019){Palit}, {Janiuk}, \&
  {Sukova}}]{2019MNRAS.487..755P}
{Palit}, I., {Janiuk}, A., \& {Sukova}, P. 2019, \mnras, 487, 755

\bibitem[{{Paxton} {et~al.}(2011){Paxton}, {Bildsten}, {Dotter}, {Herwig},
  {Lesaffre}, \& {Timmes}}]{2011ApJS..192....3P}
{Paxton}, B., {Bildsten}, L., {Dotter}, A., {et~al.} 2011, \apjs, 192, 3

\bibitem[{{Paxton} {et~al.}(2013){Paxton}, {Cantiello}, {Arras}, {Bildsten},
  {Brown}, {Dotter}, {Mankovich}, {Montgomery}, {Stello}, {Timmes}, \&
  {Townsend}}]{2013ApJS..208....4P}
{Paxton}, B., {Cantiello}, M., {Arras}, P., {et~al.} 2013, \apjs, 208, 4

\bibitem[{{Perna} {et~al.}(2014){Perna}, {Duffell}, {Cantiello}, \&
  {MacFadyen}}]{2014ApJ...781..119P}
{Perna}, R., {Duffell}, P., {Cantiello}, M., \& {MacFadyen}, A.~I. 2014, \apj,
  781, 119

\bibitem[{{Quataert} {et~al.}(2019){Quataert}, {Lecoanet}, \&
  {Coughlin}}]{2019MNRAS.485L..83Q}
{Quataert}, E., {Lecoanet}, D., \& {Coughlin}, E.~R. 2019, \mnras, 485, L83

\bibitem[{{Ram{\'\i}rez-Agudelo} {et~al.}(2013){Ram{\'\i}rez-Agudelo},
  {Sim{\'o}n-D{\'\i}az}, {Sana}, {de Koter}, {Sab{\'\i}n-Sanjul{\'\i}an}, {de
  Mink}, {Dufton}, {Gr{\"a}fener}, {Evans}, {Herrero}, {Langer}, {Lennon},
  {Ma{\'\i}z Apell{\'a}niz}, {Markova}, {Najarro}, {Puls}, {Taylor}, \&
  {Vink}}]{2013A&A...560A..29R}
{Ram{\'\i}rez-Agudelo}, O.~H., {Sim{\'o}n-D{\'\i}az}, S., {Sana}, H., {et~al.}
  2013, \aap, 560, A29

\bibitem[{{Sana} {et~al.}(2012){Sana}, {de Mink}, {de Koter}, {Langer},
  {Evans}, {Gieles}, {Gosset}, {Izzard}, {Le Bouquin}, \&
  {Schneider}}]{Sana:2012}
{Sana}, H., {de Mink}, S.~E., {de Koter}, A., {et~al.} 2012, Science, 337, 444

\bibitem[{{Schr{\o}der} {et~al.}(2020){Schr{\o}der}, {MacLeod}, {Loeb},
  {Vigna-G{\'o}mez}, \& {Mandel}}]{2020ApJ...892...13S}
{Schr{\o}der}, S.~L., {MacLeod}, M., {Loeb}, A., {Vigna-G{\'o}mez}, A., \&
  {Mandel}, I. 2020, \apj, 892, 13

\bibitem[{{Shapiro} \& {Teukolsky}(1986)}]{1986bhwd.book.....S}
{Shapiro}, S.~L., \& {Teukolsky}, S.~A. 1986, {Black Holes, White Dwarfs and
  Neutron Stars: The Physics of Compact Objects}

\bibitem[{{Smartt}(2015)}]{2015PASA...32...16S}
{Smartt}, S.~J. 2015, \pasa, 32, e016

\bibitem[{{Sukhbold} {et~al.}(2016){Sukhbold}, {Ertl}, {Woosley}, {Brown}, \&
  {Janka}}]{2016ApJ...821...38S}
{Sukhbold}, T., {Ertl}, T., {Woosley}, S.~E., {Brown}, J.~M., \& {Janka}, H.~T.
  2016, \apj, 821, 38

\bibitem[{{Sukov{\'a}} {et~al.}(2017){Sukov{\'a}}, {Charzy{\'n}ski}, \&
  {Janiuk}}]{2017MNRAS.472.4327S}
{Sukov{\'a}}, P., {Charzy{\'n}ski}, S., \& {Janiuk}, A. 2017, \mnras, 472, 4327

\bibitem[{{Sukov{\'a}} \& {Janiuk}(2015)}]{2015MNRAS.447.1565S}
{Sukov{\'a}}, P., \& {Janiuk}, A. 2015, \mnras, 447, 1565

\bibitem[{{Tchekhovskoy} {et~al.}(2011){Tchekhovskoy}, {Narayan}, \&
  {McKinney}}]{2011MNRAS.418L..79T}
{Tchekhovskoy}, A., {Narayan}, R., \& {McKinney}, J.~C. 2011, \mnras, 418, L79

\bibitem[{{Ugliano} {et~al.}(2012){Ugliano}, {Janka}, {Arcones}, \&
  {Marek}}]{2012ASPC..453...91U}
{Ugliano}, M., {Janka}, H., {Arcones}, A., \& {Marek}, A. 2012, Astronomical
  Society of the Pacific Conference Series, Vol. 453, {Explosion and Remnant
  Systematics of Neutrino-driven Supernovae for Spherically Symmetric Models},
  ed. R.~{Capuzzo-Dolcetta}, M.~{Limongi}, \& A.~{Tornamb{\`e}}, 91

\bibitem[{{Villata}(1992)}]{1992MNRAS.257..450V}
{Villata}, M. 1992, \mnras, 257, 450

\bibitem[{{Weidner} \& {Vink}(2010)}]{2010A&A...524A..98W}
{Weidner}, C., \& {Vink}, J.~S. 2010, \aap, 524, A98

\bibitem[{{Woosley}(1993)}]{1993ApJ...405..273W}
{Woosley}, S.~E. 1993, \apj, 405, 273

\bibitem[{{Woosley}(2017)}]{2017ApJ...836..244W}
---. 2017, \apj, 836, 244

\bibitem[{{Woosley} \& {Heger}(2006)}]{2006ApJ...637..914W}
{Woosley}, S.~E., \& {Heger}, A. 2006, \apj, 637, 914

\bibitem[{{Yuan} \& {Narayan}(2014)}]{2014ARA&A..52..529Y}
{Yuan}, F., \& {Narayan}, R. 2014, \araa, 52, 529

\bibitem[{{Zalamea} \& {Beloborodov}(2009)}]{Zalamea09}
{Zalamea}, I., \& {Beloborodov}, A.~M. 2009, \mnras, 398, 2005

\end{thebibliography}
\end{document}